\documentclass[11pt]{article}

\usepackage{csvsimple}
\usepackage{amsfonts}
\usepackage{amsthm}
\usepackage{booktabs}
\usepackage{amsmath}
\usepackage{multicol}
\usepackage{longtable}
\usepackage{amssymb}
\usepackage{blkarray}
\usepackage{latexsym}
\usepackage{bbm}
\usepackage[pdftex]{graphicx,color}
\usepackage[table,xcdraw]{xcolor}
\usepackage{enumerate}
\usepackage{verbatim}
\usepackage{amsfonts}
\usepackage{algorithm}
\usepackage{algorithmic}
\usepackage{epsfig}
\usepackage{subcaption}
\usepackage{tikz}
\usepackage{booktabs}
\usepackage{geometry}
\usepackage{booktabs}
\usepackage{geometry}
\usepackage{hyperref}
\usepackage{tabularx}
\usepackage{multirow}
\usepackage{bm} %bold
\usepackage{caption} % remove caption label
\usepackage{bbm, color} %(for writing indicator function)
\usepackage{amssymb,amsmath,tikz,graphicx,float,xspace,chemarrow}
\usepackage{graphicx}
\usepackage[utf8]{inputenc}
\makeatletter

\newcommand{\be}{\begin{equation}}
\newcommand{\ee}{\end{equation}}
\newcommand{\bea}{\begin{eqnarray}}
\newcommand{\eea}{\end{eqnarray}}

\newcommand{\distas}[1]{\mathbin{\overset{#1}{\kern\z@\sim}}}%
\newsavebox{\mybox}\newsavebox{\mysim}
\newcommand{\distras}[1]{%
  \savebox{\mybox}{\hbox{\kern3pt$\scriptstyle#1$\kern3pt}}%
  \savebox{\mysim}{\hbox{$\sim$}}%
  \mathbin{\overset{#1}{\kern\z@\resizebox{\wd\mybox}{\ht\mysim}{$\sim$}}}%
}

\newtheorem{theorem}{Theorem}
\newtheorem{lemma}{Lemma}

\newtheorem{remark}{Remark}

\newtheorem{definition}{Definition}
\newtheorem{proposition}{Proposition}

\definecolor{LightCyan}{rgb}{0.88,1,1}
\definecolor{shadecolor}{rgb}{0.01,0.4,.8}
\begin{document}
\begin{center} { \large \sc 
Multivariate Interval-Valued Models in Frequentist and Bayesian Schemes}

{\bf Ali Sadeghkhani}\\
Department of Mathematics and Statistics,\\ University of Windsor, ON, Canada
\\ Email: \texttt{sadeghk@uwindsor.ca}
\vspace{1cm}

{\bf Abdolnasser Sadeghkhani $^{*}$}\\
Department of Mathematics and Statistics,\\ North Carolina Agricultural and Technical State University, NC, USA
\\ E-mail: \\ $*$ Corresponding author: \texttt{asadeghkhani@ncat.edu}
\vskip 0.1in
\end{center}
\vspace*{0.5cm}

\normalsize
\begin{abstract}
In recent years, addressing the challenges posed by massive datasets has led researchers to explore aggregated data, particularly leveraging interval-valued data, akin to traditional symbolic data analysis. While much recent research, with the exception of Samdai et al. (2023) who focused on the bivariate case, has primarily concentrated on parameter estimation in single-variable scenarios, this paper extends such investigations to the multivariate domain for the first time. We derive maximum likelihood (ML) estimators for the parameters and establish their asymptotic distributions. Additionally, we pioneer a theoretical Bayesian framework, previously confined to the univariate setting, for multivariate data. We provide a detailed exposition of the proposed estimators and conduct comparative performance analyses. Finally, we validate the effectiveness of our estimators through simulations and real-world data analysis.

\end{abstract}

\noindent{\it AMS (2000) subject classification}: Primary: 62C10; Secondary: 62F15, 62E15
%%%%%%%%%%%%%%%%%%%%

\section{Introduction}

Symbolic data analysis (Diday, 1987) is a prominent field within statistical data analysis that focuses on understanding and modeling data represented in distributional form, known as symbols. These symbols can encompass various formats, including intervals, histograms, and other distributional representations. The foundational concept of symbolic data analysis is rooted in considering the symbol as the primary statistical unit of interest, necessitating inference at this level (Billard and Diday, 2012). Unlike a classical random variable taking values in $\mathbb{R}^p$, a symbolic random variable consists of a hypercube in $\mathbb{R}^p$.

\noindent Interval-valued data (a hyper rectangular in $p$ dimension), as a special case of symbolic data, provides a structured representation for information that inherently exists within intervals rather than precise point values.

\noindent Examples abound across various fields, illustrating the versatility and applicability of interval-valued data. In finance, for instance, stock prices are often depicted as intervals to accommodate market fluctuations and volatility, providing analysts with a range rather than a single price point. Similarly, environmental monitoring utilizes intervals to report measurements like temperature or pollution levels, acknowledging variations and errors inherent in the data collection process. In medical diagnostics, interval-valued data emerges prominently, especially in scenarios where test results or patient parameters exhibit uncertainty and variability. Blood pressure readings or cholesterol levels, for instance, may be communicated as intervals rather than precise values, acknowledging the inherent uncertainty in medical measurements. See e.g. Billard and Diday (2003, 2012) and Billard (2011) for an extensive, detailed overview and examples of symbolic data and their analysis, including interval-valued data.

Thanks to advancements in computational statistics, the Bayesian method effectively utilizes prior beliefs to update parameter estimates based on observed data. While the exploration of interval-valued data within the Bayesian framework is relatively novel in literature, Xu and Kin (2022) recently employed Jeffery's prior to contrast Bayesian estimates for such data. Their approach involves employing a Gibbs sampler to sample from the posterior. In this paper, we introduce priors for parameters of multivariate interval-valued model that yield results akin to those of Xu and Kin's. Importantly, our proposed priors allow for analytical determination of the posterior, eliminating the need for computational methods like the Gibbs sampler, particularly advantageous in high-dimensional scenarios (large $p$). Furthermore, this manuscript extends frequentist results presented by Samadi (2023), specifically when $p=2$.

This paper is organized as follows: Section \ref{sec-2} introduces key definitions, the likelihood function, and the maximum likelihood (ML) estimation method for $p$-variate interval-valued data, as well as some asymptotic properties. In Section \ref{Sec:Bayes}, we bring the interval-valued data into the Bayesian framework by introducing priors on the parameters, proving closed-form Bayesian estimations for the corresponding parameters, and demonstrating that the Bayesian estimators dominate the corresponding ML estimators under $L_2$ and entropy loss functions. Section \ref{sec:sim} performs extensive simulation studies and real data analysis of interval-valued data with different dimensions to evaluate the performance of the proposed methods. Finally, we present some concluding remarks in Section \ref{sec:concluding}.
\newpage
\section{Multivariate interval-valued likelihood function}\label{sec-2}

We begin this section with two definitions that will be used throughout the paper.

\begin{definition}[Wishart Distribution]

We say that $\bm{A}$ is distributed as a Wishart distribution $\mathcal{W}_p(m, \bm{V})$, where $\bm{A}$ is a $p \times p$ positive definite symmetric matrix, $m$ is the degrees of freedom, and $\bm{V}$ is the scale matrix.
The corresponding probability density function (PDF) is given by

\begin{equation}\label{Wishart}
\mathcal{W}_p(\bm{A} \mid m, \bm{V}) = \frac{{|\bm{A}|^{\frac{{m - p - 1}}{2}}}}{{2^{\frac{{m p}}{2}} |\bm{V}|^{\frac{m}{2}} \Gamma_p\left(\frac{m}{2}\right)}} \exp\left(-\frac{1}{2} \text{tr}(\bm{V}^{-1} \bm{A})\right)\,,
\end{equation}
where $|\cdot|$ denotes the determinant, $\text{tr}(\cdot)$ denotes the trace, and $\Gamma_p(\cdot)$ is the multivariate generalization of the gamma function, given as
\[\Gamma_p(z) = \pi^{\frac{p(p-1)}{4}} \prod_{i=1}^{p} \Gamma(z + \frac{1 - i}{2})
\,= \int_{\bm{A}>0} \exp \left(-\text{tr}(\bm{A})\right) \left|\bm{A}\right|^{a - \frac{p+1}{2}} d\bm{A}\,,\displaystyle \Re (a)>(p-1)/2. 
\]

Note that in Equation (\ref{Wishart}), we must have $m \geq p$ to ensure that the symmetric matrix $\bm{A}$ is invertible. Furthermore, $\mathbb{E}[\bm{A}] = m\bm{V}$.
\end{definition}
\begin{definition}[Inverse Wishart Distribution]\label{def:iw}
 
If $\bm{B} = \bm{A}^{-1}$, then $\bm{B}$ follows the inverse Wishart distribution $\mathcal{IW}_p(m, U)$, where the scale matrix is denoted as $U = \bm{V}^{-1}$, %This is written as $\mathcal{IW}(\bm{B} \mid m, \bm{U})$, 
and its PDF is given by

\begin{equation}\label{inverseWishart}
\mathcal{IW}_p(\bm{B} \mid m, \bm{U}) = \frac{{|\bm{U}|^{\frac{m}{2}}}\,|\bm{B}|^{-\frac{{m + p + 1}}{2}}}{{ 2^{\frac{{m p}}{2}} \Gamma_p\left(\frac{m}{2}\right)}} \exp\left(-\frac{1}{2} \text{tr}(\bm{U} \bm{B}^{-1})\right)\,, 
\end{equation}
with $\mathbb{E}[\bm{B}] = \bm{U}/(m-p-1)$ for $m>p+1$.
\end{definition}

%%%%%%%%%
The first step in studying interval-valued data is through descriptive statistics. Bertrand and Goupil (2000) examined the univariate random interval, considering $X_{1i}=[a_{1i}, b_{1i}]$, where $a_{1i} < b_{1i}$ for $i=1, \dots, n$, under the assumption that points are uniformly spread across the intervals. They derived the sample mean and variance as $\bar{X}_1=(2n)^{-1} \sum_{i=1}^n (a_{1i}+bi)$, and $S^2_{X_1}=(3n)^{-1}\sum_{i=1}^n (a_{1i}^2+a_{1i} b_{1i}+b_{1i}^2)-n^{-1}\bar{X}_1$.  Billard (2008) examined the sample covariance function by considering a second random variable $X_{2i}=[a_{2i}, b_{2i}]$, where $a_{2i} < b_{2i}$, resulting in
\begin{align*}
\begin{split}
S_{X_1 X_2} = &(6n)^{-1}\sum_{i=1}^n \Bigl(2(a_{1i}-\bar{X}_1)(a_{2i}-\bar{X}_2) \\
&+ (a_{1i}-\bar{X}_1)(b_{2i}-\bar{X}_2) + (b_{1i}-\bar{X}_1)(a_{2i}-\bar{X}_2) \\
&+ 2(a_{1i}-\bar{X}_1)(b_{2i}-\bar{X}_2) \Bigr)\,.
\end{split}
\end{align*}
If $a_{1i}=b_{1i}$ for each $i=1, \dots, n$, it means that each interval $X_{1i}$ collapses into a single point rather than representing a range. This is essentially equivalent to dealing with point-valued data rather than interval-valued data. Therefore, in this case, the data can be treated as point-valued rather than interval-valued. Billard (2008) and Samadi et al. (2023) expanded upon the uniform distribution assumptions proposed by Bertrand and Goupil (2000), extending the results to include triangular and Pert distributions (Clark, 1962). 

 \noindent In order to construct the multivariate likelihood function of interval-valued data based on the uniformly spread assumption, we consider $\bm{X}=(X_1, \dots, X_p)$ representing a $p$-variate random variable with interval-valued realizations $\bm{X}i=(X{1i}, \dots, X_{pi})$, where $X_{ji}=[a_{ji}, b_{ji}]$ and $a_{ji}\leq b_{ji}$ (intervals can be open or closed at either end), for $i=1, \dots, n$ and $j=1, \dots, p$, representing hyper-rectangles in $\mathbb{R}^p$. 
 
Given that each variable has aggregated observed values over an interval, it is necessary to consider the internal distribution of those values within the interval. Adapting from Le-Rademacher and Billard (2011), there exists a one-to-one correspondence between $\bm{X}_i=(X_{1i}, \dots, X_{pi})$ and $\bm{\Theta}=(\bm{\Theta}_1, \bm{\Theta}_2)^\top$, where $\bm{\Theta}_{1}$ and $\bm{\Theta}_{2}$ represent the mean and the variance-covariance matrix of the internal distribution and can be obtained by
\begin{align}
\bm{\Theta}_{i1}\,=&\frac{1}{2}\left(a_{1i}+b_{1i}, \dots, a_{pi}+b_{pi}\right)^\top \,,\label{thetai1}\\
\bm{\Theta}_{i2}\,=&\frac{1}{12}\text{diag}((b_{1i}-a_{1i})^2, \dots, (b_{pi}-a_{1i})^2)+\frac{1}{12}\sum_{j \neq k} (b_{ji}-a_{ji})(b_{ki}-a_{ki}) \,\label{thetai2}\,.
\end{align}
It is worth mentioning that equation (\ref{thetai2}) represents a matrix with diagonal elements \( \Theta^{x_1}_{i2}, \Theta^{x_2}_{i2}, \dots, \Theta^{x_p}_{i2} \) and off-diagonal elements \( \Theta^{x_j x_k}_{i2} \) for \( j \neq k \).

Since $\bm{X}_i$ is a random variable, the corresponding parameter $\bm{\Theta}$ varies and takes different values.
Suppose the PDF of $\bm{X}_i$, denoted by $f_i^{\bm{X}_i}( \bm{x}_i; \bm{\Theta})$ and consequently can be expressed as a joint density of $\bm{\Theta}=(\bm{\Theta}_1, \bm{\Theta}_2)^\top$ given by
%%%%%%%%%%%%%%%
\begin{align}
\bm{\Theta}_{i1}\,&=\,(\Theta^{x_1}_{i1}, \dots, \Theta^{x_p}_{i1} )^\top \sim \mathcal{N}_p(\bm{\mu}, \bm{\Sigma}) \,, \label{model1}\\
\bm{\Theta}_{i2}\,&=\,\begin{pmatrix}
\Theta^{x_1}_{i2} & \Theta^{x_1 x_2}_{i2} & \cdots & \Theta^{x_1 x_{p-1}}_{i2} & \Theta^{x_1 x_p}_{i2} \\
\Theta^{x_2 x_1}_{i2} & \Theta^{x_2}_{i2} & \cdots & \Theta^{x_2 x_{p-1}}_{i2} & \Theta^{x_2 x_p}_{i2} \\
\vdots & \vdots & \ddots & \vdots & \vdots \\
\Theta^{x_{p-1} x_1}_{i2} & \Theta^{x_{p-1} x_2}_{i2} & \cdots & \Theta^{x_{p-1}}_{i2} & \Theta^{x_{p-1} x_p}_{i2} \\
\Theta^{x_p x_1}_{i2} & \Theta^{x_p x_2}_{i2} & \cdots & \Theta^{x_p x_{p-1}}_{i2} & \Theta^{x_p}_{i2} \\
\end{pmatrix}\sim \mathcal{W}_p(m, \bm{\Lambda})\,. \label{model2}
\end{align}
Next, we establish the likelihood function based on intervals.
%%%%%%%%%%%%%%%%%%%%%%%%%%%%%%%%%

\subsection{ML estimators and related properties}
Let $\bm{S} = \sum_{i=1}^{n} (\bm{\Theta}_{i1} - \bar{\bm{\Theta}}_1)(\bm{\Theta}_{i1} - \bar{\bm{\Theta}}_1)^\top$, where $\bar{\bm{\Theta}}$ is the mean vector of parameters $\bm{\Theta}_{i1}$ for $i = 1, \dots, n$, given by $\bar{\bm{\Theta}}_1 = (\bar{\Theta}_1, \dots, \bar{\Theta}_n)^\top$. 

Provided that $\{\bm{\Theta}_{i1}\}_{i=1}^n$ are independent and identically distributed (iid) and independent of iid $\{\bm{\Theta}_{i2}\}_{i=1}^n$, then the likelihood functions from equations (\ref{model1}) and (\ref{model2}) are given respectively by
\begin{align}
    L_1\,=&\,L_1(\bm{\mu}, \bm{\Sigma} \mid \bm{\Theta}_{11}, \dots, \bm{\Theta}_{n1})\,=\,|\bm{\Sigma}|^{-\frac{n}{2}}\exp\left(-\frac{1}{2} \sum_{i=1}^n (\bm{\theta}{i1} -\bm{\mu})^\top \bm{\Sigma}^{-1} (\bm{\theta}{i1} -\bm{\mu})\right) \nonumber \\
    =&\,|\bm{\Sigma}|^{-\frac{n}{2}}    
    \exp\left(-\frac{1}{2} \text{tr}(\bm{S} \bm{\Sigma}^{-1})-\frac{n}{2}  (\bar{\bm{\theta}}_1 -\bm{\mu})^\top (\bar{\bm{\theta}}_1 - \bm{\mu})\right)\,, \label{L1}\\
    L_2\,=&\,L_2(\bm{\Lambda} \mid \bm{\Theta}_{12}, \dots, \bm{\Theta}_{n2})\,=\,|\bm{\Lambda}|^{-\frac{nm}{2}}\exp\left(-\frac{1}{2} \text{tr}(\sum_{i=1}^n \bm{\theta}_{i2} \bm{\Delta}^{-1})\right)\,.\label{L2}
\end{align}
The ML estimators of unknown vector $\bm{\mu}$, and matrices $\bm{\Sigma}$ and $\bm{\Lambda}$ are presented in the following theorem.

\begin{theorem}\label{MLest}
    The ML estimators of parameters $\bm{\mu}$, and matrices $\bm{\Sigma}$ and $\bm{\Lambda}$ are given by
    \begin{align}
\hat{\bm{\mu}}^{ML}\,=\,&\bar{\bm{\Theta}}_1\,,\label{muML}\\
\hat{\bm{\Sigma}}^{ML}\,=\,&\frac{\bm{S}}{n}\,,\label{SigML}\\ 
\hat{\bm{\Lambda}}^{ML}\,=\,&\frac{\sum_{i=1}^n \bm{\Theta}_{i2}}{nm}\,.\label{lamML}
    \end{align}
\end{theorem}
\begin{proof}
Considering the likelihood function \( L_1 \) in (\ref{L1}), and taking the derivative of \( L_1 \) with respect to \( \bm{\mu} \) and setting it to zero gives
\begin{align*}
\frac{\partial L_1}{\partial \bm{\mu}} &= \frac{\partial}{\partial \bm{\mu}} \left( -\frac{n}{2}  (\bar{\bm{\theta}}_1 -\bm{\mu})^\top (\bar{\bm{\theta}}_1 - \bm{\mu}) \right) \\
&= -\frac{n}{2}  \frac{\partial}{\partial \bm{\mu}} \left( \bar{\bm{\theta}}_1^\top \bar{\bm{\theta}}_1 - 2\bar{\bm{\theta}}_1^\top \bm{\mu} + \bm{\mu}^\top \bm{\mu} \right) \\
%&= -\frac{n}{2} \left( -2\bar{\bm{\theta}}_1 + 2\bm{\mu} \right) \\
&= n(\bar{\bm{\theta}}_1 - \bm{\mu}) = 0\,,
\end{align*}

and solving for \( \bm{\mu} \) gives equation (\ref{muML}).

Analogously, by taking the derivative of \( L_1 \) with respect to \( \bm{\Sigma} \) and setting it to zero, we have
\begin{align*}
\frac{\partial L_1}{\partial \bm{\Sigma}} &= \frac{\partial}{\partial \bm{\Sigma}} \left( -\frac{1}{2} \text{tr}(\bm{S} \bm{\Sigma}^{-1}) \right) \\
&= -\frac{1}{2} \frac{\partial}{\partial \bm{\Sigma}} \left( \text{tr}(\bm{S} \bm{\Sigma}^{-1}) \right) = 0\,,
\end{align*}
solving for \( \bm{\Sigma} \) gives
equation (\ref{SigML}).

In order to find the ML estimator $\bm{\Lambda}$, having the likelihood function \( L_2 \) in (\ref{L2})
and taking derivative of it with respect to \( \bm{\Lambda} \) and setting it to zero gives
\begin{align*}
\frac{\partial L_2}{\partial \bm{\Lambda}} &= -\frac{nm}{2} |\bm{\Lambda}|^{-\frac{nm}{2}-1} \exp\left(-\frac{1}{2} \text{tr}\left(\sum_{i=1}^n \bm{\theta}_{i2} \bm{\Delta}^{-1}\right)\right) = 0
\end{align*}

Solving for \( \bm{\Lambda} \) yields equation (\ref{lamML}). This completes the proof.

\end{proof}

\subsection{Asymptotic properties of ML estimators}
\begin{theorem}
    Consider \( \text{Sym}(p) \), the set of \( p \times p \) real symmetric matrices, and let \( \mathbb{P}(p) \subseteq \text{Sym}(p) \) represent the subset consisting of symmetric positive-definite matrices that forms a convex regular cone. Setting $\bm{\omega}=(\bm{\mu}, \bm{\Sigma}) \in \Omega=\mathbb{R}^p \times \mathbb{P}(p)$, then
    \begin{align*}
        \sqrt{n}(\hat{\bm{\omega}}-\bm{\omega}) \xrightarrow{d} \mathcal{N}_m(\bm{0}_m, I^{-1}(\bm{\omega}))\,,
    \end{align*}
with \begin{align*}
  I_{ij}(\bm{\omega}) \,=\, \left[\frac{\partial \bm{\mu}}{\partial \omega_i}\right]^\top \bm{\Sigma}^{-1} \frac{\partial \bm{\mu}}{\partial \omega_j} + \frac{1}{2} \operatorname{tr} \left( \bm{\Sigma}^{-1} \frac{\partial \bm{\mu}}{\partial \omega_i} \bm{\Sigma}^{-1} \frac{\partial \mu}{\partial \omega_j} \right)\,,
\end{align*}
and $m=dim(\Omega)=p(p+3)/2$.
\end{theorem}
\begin{proof}
   The symmetric semi-positive fisher information matrix (eg., Amari 2016) is given by $I(\bm{\omega})=\mathbb{V}[\nabla \log  \mathcal{N}_m(\Theta_1 \mid \bm{\mu}, \bm{\Sigma})]$, where is a PDF of $p$-varaite normal with mean vector $\mathbb{E}[\Theta]=\bm{\mu}$, $\mathbb{V}[\Theta]=\bm{\Sigma}$, and $\mathbb{V}(\cdot)$ is the variance-covariance matrix.
 As discussed in Nielsen (2023), the fisher information matrix can be written as follows
 \begin{align*}
I(\bm{\omega}) &= \text{Cov}\left[\nabla \log \mathcal{N}_m(\Theta_1 \mid \bm{\mu}, \bm{\Sigma}), \bm{\Sigma})\right] \\
&= \mathbb{E} \left[ h \nabla \log \mathcal{N}_m(\Theta_1 \mid \bm{\mu}, \bm{\Sigma})) \nabla \log \mathcal{N}_m(\Theta_1 \mid \bm{\mu}, \bm{\Sigma}))^\top \right] \\
&= -\mathbb{E} \left[ \nabla^2 \log \mathcal{N}_m(\Theta_1 \mid \bm{\mu}, \bm{\Sigma})) \right]
\end{align*}
  For multivariate distributions parameterized by an \( m \)-dimensional vector \( \bm{\psi} = (\psi_1, \ldots, \psi_p, \psi_{p+1}, \ldots, \psi_m) \in \mathbb{R}^m \), with \( \bm{\mu} = (\psi_1, \ldots, \psi_p) \) and \( \bm{\Sigma}(\bm{\psi}) = \text{vech}(\psi_{p+1}, \ldots, \psi_m) \), where \( \text{vech}(\cdot) \) refers to the vech operator. Then we have $I(\bm{\omega})\,=\,\left[I_{ij}(\bm{\omega})\right]$,  with
\begin{align*}
  I_{ij}(\bm{\omega}) \,=\, \left[\frac{\partial \bm{\mu}}{\partial \omega_i}\right]^\top \bm{\Sigma}^{-1} \frac{\partial \bm{\mu}}{\partial \omega_j} + \frac{1}{2} \operatorname{tr} \left( \bm{\Sigma}^{-1} \frac{\partial \bm{\mu}}{\partial \omega_i} \bm{\Sigma}^{-1} \frac{\partial \bm{\mu}}{\partial \omega_j} \right).
\end{align*}
see Skovgaard (1984), and Barachant (2013) for more information.
\end{proof}
%%%%%%%
\begin{proposition}
It can be checked that for $p=1$ and $p=2$, the Fisher information matrices are simplified to the following.
\begin{align*}
I(\bm{\omega})\,=\,&\begin{bmatrix}
1/\sigma^2 & 0 \\
0 & 1/\sigma^4 \\
\end{bmatrix} \,,\,\, \text{\,with\,\,\,\,\,\,} \bm{\omega}=(\mu, \sigma^2)\,, \\
I(\bm{\omega})\,=\,&
\begin{bmatrix}
 \bm{A}_{2\times2} & \bm{0}_{2\times3} \\
\bm{0}_{3\times2} & \bm{B}_{2\times2} \\
\end{bmatrix}\,,\,\, \text{\,with\,\,\,\,\,\,} \bm{\omega}=(\mu_1, \mu_2, \sigma_1^2, \sigma_2^2, \rho)\,, 
\end{align*}
and symmetric matrices are given by
\begin{align*}
    \bm{A}\,=\,\begin{bmatrix}
\frac{-1}{(1-\rho^2)\sigma_1^2} & \frac{\rho}{(1-\rho^2)\sigma_1 \sigma_2} \\
 \frac{\rho}{(1-\rho^2)\sigma_1 \sigma_2}& \frac{1}{(\rho^2-1)\sigma_2^2} \\
\end{bmatrix}\,,\,\,\,   \bm{B}\,=\,\begin{bmatrix}
\frac{-\rho^2+1}{(\rho^2-1)^2} & \frac{\rho}{ (1-\rho^2)\sigma_1} & \frac{\rho}{ (1-\rho^2)\sigma_2} \\
 \frac{\rho}{ (1-\rho^2)\sigma_1}& \frac{2-\rho^2}{(\rho^2-1)\sigma_1^2}&\frac{\rho^2}{(\rho^2-1)\sigma_1\sigma_2} \\
 \frac{\rho}{ (1-\rho^2)\sigma_2} & \frac{\rho^2}{ (1-\rho^2)\sigma_1 \sigma_2} & \frac{2-\rho^2}{(\rho^2-1)\sigma_2^2}
\end{bmatrix}\,.
\end{align*}
\end{proposition}
%%%%%

\section{Bayesian set up}\label{Sec:Bayes}

We begin by proposing prior distributions on the parameters and subsequently derive the posterior distribution based on the given likelihood functions (\ref{L1}) and (\ref{L2}) corresponding to models (\ref{model1}) and (\ref{model2}), respectively.

Consider the following priors on parameters $\bm{\mu}, \bm{\Sigma}$ in (\ref{model1}) and $\bm{\Lambda}$ in (\ref{model2}) as below
\begin{align}
    \pi(\bm{\mu}, \bm{\Sigma}) &\propto |\bm{\Sigma}|^{-\frac{p+2}{2}},\label{prior1}\\
    \pi(\bm{\Lambda}) &\propto |\bm{\Lambda}|^{-\frac{p+1}{2}}.\label{prior2}
\end{align}

Let $\bm{\Theta} = \left\{(\bm{\Theta}_{i1}, \bm{\Theta}_{i2})^\top\right\}_{i=1}^n$, and suppose that priors (\ref{prior1}) and (\ref{prior2}) are independent. Thus, we have
\begin{equation}\label{prios}
    \pi(\bm{\mu}, \bm{\Sigma}, \bm{\Lambda}) \propto |\bm{\Sigma}|^{-\frac{p+2}{2}} |\bm{\Lambda}|^{-\frac{p+1}{2}}.
\end{equation}

The posterior distribution of $\bm{\mu}, \bm{\Sigma}, \bm{\Lambda}$ can then be written as
\begin{align}
    \pi(\bm{\mu}, \bm{\Sigma}, \bm{\Lambda} \mid \bm{\Theta}) &\propto L_1(\bm{\mu}, \bm{\Sigma} \mid \bm{\Theta}_{11}, \dots, \bm{\Theta}_{1n}) \,L_2(\bm{\Lambda} \mid \bm{\Theta}_{12}, \dots, \bm{\Theta}_{n2}) \,\pi(\bm{\mu}, \bm{\Sigma}, \bm{\Lambda}) \nonumber\\
  &\propto  |\bm{\Sigma}|^{-\frac{n+p+2}{2}} |\bm{\Lambda}|^{-\frac{nm+p+2}{2}} \exp\left(-\frac{1}{2} \text{tr}(\bm{S} \bm{\Sigma}^{-1})-\frac{n}{2}\sum_{i=1}^n  (\bm{\theta}_{i1} -\bm{\mu})^\top \bm{\Sigma}^{-1}(\bm{\theta}_{i1} - \bm{\mu})\right) \nonumber \\
 &\times \exp\left(-\frac{1}{2} \text{tr}(\sum_{i=1}^n \bm{\theta}_{i2} \bm{\Lambda}^{-1})\right).\label{postprop}
\end{align}

The following lemma provides the full conditional posterior distributions associated with the posterior distribution in (\ref{postprop}).

\begin{lemma}\label{fullcond} 
Full conditional distributions
 associated with the posterior distribution in (\ref{postprop}) are given by
\begin{align}
   \bm{\mu} \mid \bm{\Sigma}, \bm{\Theta}_1\, &\sim\, \mathcal{N}_p(\bar{\bm{\Theta}}_1, \bm{\Sigma}/n)\,, \label{fullmu} \\
   \bm{\Sigma} \mid \bm{\mu}, \bm{\Theta}_1\, &\sim\, \mathcal{IW}_p(n+1, (\bar{\bm{\Theta}}_{1} -\bm{\mu}) (\bar{\bm{\Theta}}_{1} -\bm{\mu})^\top/n)\,, \label{fullsigma} \\
   \bm{\Lambda} \mid \bm{\Theta}_2\, 
   &\sim\, \mathcal{IW}_p(nm, \sum_{i=1}^n \bm{\Theta}_{i2})\,.\label{fulllambda}
\end{align}
\end{lemma}
\begin{proof}
    The proof is straightforward and hence is omitted.
\end{proof}
\subsection{Loss functions}
The most common loss function for estimating vector $\bm{\mu}$ using $\hat{\bm{\mu}}$ is $L_2$ loss, $|| \bm{\mu}-\hat{\mu}||^2$, while the common loss function in the matrix form is the entropy loss (Stein 1956)
\begin{align}\label{ent-loss}
    \mathcal{L}(\bm{B},  \hat{\bm{B}})\,=\,\text{tr} \left(\hat{\bm{B}} \bm{B}^{-1}\right)-\log
    \begin{vmatrix} \hat{\bm{B}} \bm{B}^{-1}\end{vmatrix}-p\,,
\end{align}
where $B$ is a $p \times p$ symmetric matrix. The Bayesian estimator for the matrix estimator is the posterior mean.

The corresponding risk function to loss function (\ref{ent-loss}) is given by
\begin{align}
    \label{risk}
    \mathcal{R}(\hat{\bm{B}}, \bm{B})\,=\,\mathbb{E}\,\left[\mathcal{L}(\hat{\bm{B}}, \bm{B})\right]\,.
\end{align}

\begin{theorem} \label{Bayesests}
   Consider model \ref{model1}, \ref{model2}, and the prior (\ref{prios}). The The Bayes estimators of parameters $\bm{\mu}$ (with respect to $L_2$ loss), $\bm{\Sigma}$, and $\bm{\Lambda}$ (with respect to entropy loss function in \ref{ent-loss}) are given by
   \begin{align}
\hat{\bm{\mu}}\,=\,&\bar{\bm{\Theta}}_1\,,\\
\hat{\bm{\Sigma}}\,=\,&\frac{\bm{S}}{n-p}\,, \label{SigBayes}\\ 
\hat{\bm{\Lambda}}\,=\,&\frac{\sum_{i=1}^n \bm{\Theta}_{i2}}{nm-p-1}\,.\label{LamBayes}
\end{align}
\end{theorem}

\begin{proof}
    Given that Bayes estimators are the expectations of corresponding marginal distributions, and with the posterior distributions available in (\ref{postprop}), we integrate over $\bm{\Sigma}$ and $\bm{\Lambda}$, $\bm{\mu}$ and $\bm{\Lambda}$, and eventually over $\bm{\Sigma}$ and $\bm{\mu}$, yielding
    \begin{align}
     \bm{\mu} \mid \bm{\Theta}  &\sim \mathcal{T}_p(\bar{\bm{\Theta}}_1, \frac{\bm{S}}{n+p+1}, n+1-p)\,, \label{marmu} \\
     \bm{\Sigma} \mid \bm{\Theta}  &\sim  \mathcal{IW}_p(n+1, \bm{S})\,,
     %\label{marsigma}
     \nonumber\\
     \bm{\Lambda} \mid  \bm{\Theta}  &\sim \mathcal{IW}_p(nm, \sum_{i=1}^n \bm{\Theta}_{i2})\,,
     %\label{marlambda} 
     \nonumber
    \end{align}

where $\mathcal{T}_p(\boldsymbol{m}, \boldsymbol{A}, \nu)$ in (\ref{marmu}) represents a multivariate Student's t-distribution with mean vector $\bm{m}$, variance matrix $\bm{A}$, and $\nu$ degrees of freedom with an expectation of $\bm{m}$. Completing the proof involves using the expectation of the inverse Wishart distribution, as given in Definition \ref{def:iw}.
\end{proof}
%%%%%%%%%%%%%%%%%%%%%%
%Dom result
\begin{theorem}
Under the assumptions of Theorem \ref{Bayesests}, the Bayes estimators of the parameters $\bm{\Sigma}$ and $\bm{\Lambda}$ obtained in equations (\ref{SigBayes}) and (\ref{LamBayes}) dominate the ML estimators (\ref{SigML}) and (\ref{lamML}) obtained in Theorem \ref{MLest} under entropy loss function (\ref{ent-loss}).
\end{theorem}
\begin{proof}
Let $\mathrm{\Delta_{\Sigma}}$ and $\mathrm{\Delta_{\Lambda}}$ denote the difference in risk functions of the Bayes and ML estimators for $\bm{\Sigma}$ and $\bm{\Lambda}$, respectively. It can be easily seen that

\begin{align}
\mathrm{\Delta_{\Sigma}} &= \mathbb{E}\left[\mathcal{L}(\hat{\bm{\Sigma}}^{ML}, \bm{\Sigma})-\mathcal{L}(\hat{\bm{\Sigma}}, \bm{\Sigma})\right] = \log \frac{n}{n-1}\,. \label{r1}
\end{align}

Similarly, one can show that

\begin{align}
\mathrm{\Delta_{\Lambda}} &= \frac{nm}{nm-p-1}\,. \label{r2}
\end{align}

Both equations (\ref{r1}) and (\ref{r2}) confirm that the difference in risk functions is positive. This completes the proof.
\end{proof}

%%%%%%%%%%%%%%%%%%%%%%%%%%%
\subsection{Special cases}
\subsubsection{Univariate case}
When $p=1$, equations (\ref{model1}) and (\ref{model2}) imply that $\{\Theta_{i1}\}_{i=1}^n$ are IID from $\mathcal{N}(\mu, \sigma^2)$ and are independent of $\{\Theta_{i2}\}_{i=1}^n$, which are IID from an exponential distribution $\mathcal{E}(\lambda)$ (equivalently $\mathcal{W}_1(2, 2 \lambda)$). Furthermore, as shown by Le-Rademacher and Billard (2011), the ML estimators of parameters $\mu$, $\sigma^2$, and $\lambda$ are given by

\begin{align*}
\hat{\mu}^{ML}\,=\,\bar{\Theta}_1\,,\,\,\,\,\,\,\,\,\,\hat{\sigma}^{2ML}\,=\,\sum_{i=1}^n(\Theta_{i1}-\bar{\Theta}_1)^2/n \,\,\,\,\,\hat{\lambda}^{ML}=\,\,\frac{\sum_{i=1}^n\Theta_{i2}}{n}
\end{align*}

The corresponding posterior distribution (\ref{postprop}) in this case is $\pi(\mu, \sigma, \lambda) \propto \sigma^{-3}\lambda^{-1}$, which is also Jeffrey's prior studied by Xu and Qin (2022). Therefore, the conditional posterior distributions from equations (\ref{fullmu}), (\ref{fullsigma}), and (\ref{fulllambda}) are reduced respectively to the following
\begin{align*}
    \mu \mid \Theta, \sigma^2, \gamma \,\sim\, &\mathcal{N}(\bar{\Theta}_1, \sigma^2/n)\,, \\
    \sigma^2 \,\mid\, \Theta, \mu, \lambda \sim& \mathcal{IG}(\frac{n+1}{2}, \sum_{i=1}^n (\Theta_{i2}-\mu)^2/2)\,,\\
    \lambda \,\mid\, \Theta, \mu, \sigma^2 \sim& \mathcal{IG}(n, \sum_{i=2}^n \Theta_{i2} )\,.
\end{align*}
Moreover, the Bayes estimators can also be retrieved from Theorem (\ref{Bayesests}), with $p=1$ as below

\begin{align*}
 \hat{\mu}\,=\,\bar{\Theta}_1\,,\,\,\,\,
    \hat{\sigma}^2\,=\,\frac{\sum_{i=1}^n (\Theta_{i1}-\bar{\Theta}_1)^2}{n-1}\,,\,\,\,\,
    \hat{\lambda}\,=\, \frac{\sum_{i=1}^n\Theta_{i2}}{n}\,. 
\end{align*}
Unlike Xu and Qin (2022), we have proposed closed-form Bayesian estimators for the parameters, making the Gibbs sampler method they employed unnecessary.
\subsubsection{Bivariate case}
In this case $\{\bm{\Theta}^{x_1, x_2}_{i1}\}_{i=1}^n$ are iid from $\mathcal{N}_2(\mu_1, \mu_2, \sigma_1^2, \sigma_2^2, \rho)$ (which is corresponding $\bm{\mu}=(\mu_1, \mu_2)^\top$, and $\bm{\Sigma}= \left( \begin{smallmatrix} \sigma_1^2 & \rho \sigma_1 \sigma_2 \\ \rho \sigma_1 \sigma_2 & \sigma_2^2  \end{smallmatrix} \right)
$ in \ref{model1}) is independent of $\{\bm{\Theta}^{x_1, x_2}_{i2}\}_{i=1}^n$ are iid from $\mathcal{W}_2(m, \bm{\Lambda}= \left( \begin{smallmatrix} \lambda_{11} & \lambda_{12} \\ \lambda_{12} & \lambda_{22}  \end{smallmatrix} \right))$. 

Therefore the conditional posterior distributions are obtained using (\ref{fullmu}), (\ref{fullsigma}) and (\ref{fulllambda}) with $p=2$.
In order to obtain the ML estimators of the parameters, lets rewrite the likelihood functions (\ref{L1}) and (\ref{L2}) in this case as follows.
\begin{align}
    L_1\,=&\,(2 \pi \sigma_1 \sigma_2)^{-n} exp\left(\sum_{i=1^n}(\theta^{x_1}_{i1}-\mu_1)^2/\sigma_1^2+\theta^{x_2}_{i1}-\mu_2)^2/\sigma_2^2\right)-2\frac{\rho}{\sigma_1 \sigma_2} \theta^{x_1}_{i1}-\mu_1)\theta^{x_2}_{i1}-\mu_2)/\sigma_2^2\,,\label{L1p2}\\
    L_2\,=&\,\frac{(\lambda_{11} \lambda_{22}-\lambda_{12})^{2n} 2^{-mn} \pi^{-n/2}}{ (\Gamma(m/2) \Gamma((m-1)/2))^n}
    \prod_{i=1}^n (\theta^{x_1}_{i2} \theta^{x_2}_{i2}-\theta^{x_1 x2}_{i2})^{m/2-1} \nonumber 
    \\
    &\exp\left(-\frac{\lambda_{11} \lambda_{22}}{(\lambda_{11} \lambda_{22}-\lambda_{12}^2)} \left(\frac{\sum_{i=1}^n\theta^{x_1}_{i2}}{\lambda_{11}}+\frac{\sum_{i=1}^n\theta^{x_2}_{i2}}{\lambda_{22}}-\frac{2 \lambda_{12}}{\lambda_{11} \lambda_{22}} \sum_{i=1}^n  \theta^{x_1 x2}_{i2}\right)\right)\,. \label{l2p2}
\end{align}

Taking derivatives with respect to parameters of $\mu_i$, $\sigma_i$ for $i=1,2$, and $\rho$ from log of likelihood functions (\ref{L1p2}) and (\ref{l2p2}) (see Samadi et al. for details) results to following ML estimators.

\begin{align*}
\hat{\mu}^{ML}_i\,&=\,\bar{\Theta}^{x_i}_i\,, \,\,\,\,\hat{\sigma}^{2ML}_i\,=\,\sum_{i=1}^n(\Theta^{x_1}_{i1}-\mu_i)^2/n\,, \text{\,for\,\,\,\,\,\,}i=1,2\,,
\\
\hat{\rho}^{ML}\,&=\,\frac{\sum_{i=1}^n(\Theta^{x_i}_{i1}-\mu_1)(\Theta^{x_2}_{i1}-\mu_2)}{\sqrt{\sum_{i=1}^n(\Theta^{x_1}_{i1}-\mu_1)^2 \sum_{i=1}^n(\Theta^{x_1}_{i1}-\mu_2)^2}}\,,
\\
\hat{\lambda}_{ii}^{ML}\,&=\,\frac{\sum_{i=1}^n\Theta^{x_i}_{i2}}{nm}\,,\,\,\,\text{\,for\,\,\,\,\,\,}i=1,2\,,\hat{\lambda}_{12}^{ML}\,=\,\frac{\sum_{i=1}^n\Theta^{x_1 x_2}_{i2}}{nm}\,, 
\end{align*}
which are corresponding to Theorem \ref{Bayesests} with $p=2$.

\section{Simulation results} \label{sec:sim}
In the simulation, three scenarios are considered to generate samples of $n$ from the random variables \(\bm{\Theta}_{i1}\) and \(\bm{\Theta}_{i2}\), as described in equations (\ref{model1}) and (\ref{model2}). In the first scenario (I), samples are generated from univariate distributions, where each random variable \(\bm{\Theta}_{i1}\) and \(\bm{\Theta}_{i2}\) was independently sampled from $\mathcal{N}(\mu=2, \sigma^2=5)$, and $\mathcal{E}(\lambda=2)$, respectively. 

In the second scenario (II), bivariate distributions were employed, generating samples where the random variables \(\bm{\Theta}_{i1}\) and \(\bm{\Theta}_{i2}\) are independently sampled from $\mathcal{N}_2(\left(\begin{smallmatrix}
  2 \\
  4
\end{smallmatrix}\right), \left(\begin{smallmatrix}
  4 & 3 \\
  3 & 9\\
\end{smallmatrix}\right) )$, and $\mathcal{W}_2(m=3, \bm{\Lambda}= \left( \begin{smallmatrix} 2 & 1 \\ 1 & 5  \end{smallmatrix} \right))$. 
Finally, in the third scenario (III), trivariate distributions were utilized, resulting in samples where the random variables \(\bm{\Theta}_{i1}\) and \(\bm{\Theta}_{i2}\) are sampled from $\mathcal{N}_3(\bm{\mu}, \bm{\Sigma})$, and $\mathcal{W}_3(m=3, \bm{\Lambda})$, with
$$\bm{\mu}\,=\,\,\begin{bmatrix}
  2 \\
  4\\
6
\end{bmatrix}\,,\, \bm{\Sigma}\,=\,\begin{bmatrix}
1 & 1.4 & 0.6 \\
1.4 & 4 & 1.5 \\
0.6 & 1.5 & 9 \\
\end{bmatrix}\,,\,\,\, \bm{\Lambda}\,=\,\begin{bmatrix}
2 & 1 & 1 \\
1 & 5 & 2 \\
1 & 2 & 3 \\
\end{bmatrix}\,.$$
These scenarios allow us to compare behavior and performance of the proposed Bayesian and ML parameter estimators for different dimensions.

For each simulation within each scenario, sample sizes of $n=25, 50, 200$, and $500$, with $10,000$ iterations, are conducted. The estimated parameters using Theorems \ref{MLest} and \ref{Bayesests}, associated with the ML and Bayes estimators, are tabulated in Tables \ref{sim1}, \ref{sim2}, and \ref{sim3}, corresponding to dimensions $p=1$, $p=2$, and $p=3$ (scenarios I, II, and III).

\noindent According to Table \ref{sim1}  both methods yield similar estimates for $\mu$ across sample sizes, while Bayesian estimation tends to produce slightly higher estimates for $\sigma^2$ and $\lambda$ compared to MLE, with standard deviations also presented.

\noindent In Table \ref{sim2}, representing a simulation with $p=2$, both ML and Bayesian estimations exhibit consistency across various parameters and sample sizes. Analogous to Table \ref{sim1}, both methods yield similar estimates for $\mu_1$ and $\mu_2$, irrespective of sample size, with consistent standard deviations. However, for $\sigma^2_1$, $\sigma^2_2$, $\sigma_{12}$, $\lambda_{11}$, $\lambda_{22}$, and $\lambda_{12}$, Bayesian estimation tends to produce slightly higher estimates compared to ML, accompanied by corresponding standard deviations. This trend persists across different sample sizes, highlighting the robustness of Bayesian estimation in this scenario.

On the other hand, in Table \ref{sim3}, reflecting scenario III, we observe similar trends to those seen in simulation II (Table \ref{sim2}). Bayesian estimation consistently yields slightly higher parameter estimates compared to ML across various parameters and sample sizes.

%%%%%
\begin{table}[H]
\centering
\caption{Comparison of maximum likelihood and Bayesian estimations in Simulation 1.}
\label{sim1}
\begin{tabular}{@{}lccc@{}}
\toprule
\textbf{Parameter} & \textbf{MLE} & \textbf{Bayesian}  \\ \midrule
\multicolumn{4}{l}{\textbf{n = 25}} \\ \midrule
$\mu$  & 2.006555 (SD: 0.4395082)  & 2.006555 (SD: 0.4395082) \\
$\sigma^2$     & 4.824011 (SD: 1.396871)  & 5.025011 (SD: 1.455074)
\\
$\lambda$   &2.006947 (SD: 0.3253547)  & 2.061932 (SD: 0.3342685)  \\ \midrule
\multicolumn{4}{l}{\textbf{n = 50}} \\ \midrule
$\mu$      &2.000126 (SD: 0.3111599) &2.000126 (SD: 0.3111599)   \\
$\sigma^2$      & 4.912915 (SD: 0.9940378)
  &5.013179 (SD: 1.014324)  \\
$\lambda$    &2.001979 (SD: 0.2297495)  &  2.029032 (SD: 0.2328543)
 \\ \midrule
\multicolumn{4}{l}{\textbf{n = 200}} \\ \midrule
$\mu$       & 1.997678 (SD: 0.1580845) &1.997678 (SD: 0.1580845)  &  \\
$\sigma^2$     &4.973424 (SD: 0.4938952) &4.998416 (SD: 0.4963771)   \\
$\lambda$     & 1.999907 (SD: 0.1159792)  &2.006596 (SD: 0.116367)  \\ \midrule
\multicolumn{4}{l}{\textbf{n = 500}} \\ \midrule
$\mu$     &1.998748 (SD: 0.09948847)
  &1.998748 (SD: 0.09948847)
  \\
$\sigma^2$   & 4.990833 (SD: 0.3125381) &  .000835 (SD: 0.3131644) \\
$\lambda$    & 2.000496 (SD: 0.07324693)
  &2.003167 (SD: 0.07334472)  \\
\bottomrule
\end{tabular}
\end{table}

%%%%%%
%%%%%%
\begin{table}[H]
\centering
\caption{Comparison of maximum likelihood and Bayesian estimations in Simulation 2.}
\label{sim2}
\begin{tabular}{@{}lccc@{}}
\toprule
\textbf{Parameter} & \textbf{MLE} & \textbf{Bayesian} \\ \midrule
\multicolumn{4}{l}{\textbf{n = 25}} \\ \midrule
$\mu_1$       &1.998343 (SD: 0.3971383)  & 1.998343 (SD: 0.3971383)               \\
$\mu_2$       &3.998895 (SD: 0.5914762)   &  3.998895 (SD: 0.5914762)                \\

$\sigma^2_1$  & 3.854855 (SD: 1.116438) & 4.19006 (SD: 1.213519)      \\
$\sigma^2_2$  & 8.67281 (SD: 2.542278)     &        9.426968 (SD: 2.763346)    \\
$\sigma_{12}$  & 2.898315 (SD: 1.320275)    &    3.150342 (SD: 1.435081)    \\

$\lambda_{11}$   & 2.001881 (SD: 0.324576)     & 2.085292 (SD: 0.3381)       \\
$\lambda_{22}$   &   4.989752 (SD: 0.811227)     &  5.197658 (SD: 0.8450281)        \\
$\lambda_{12}$   &   1.003482 (SD: 0.3770854)     &  1.045294 (SD: 0.3927973)               \\
\midrule
\multicolumn{4}{l}{\textbf{n = 50}} \\ \midrule
$\mu_1$  &    1.997207 (SD: 0.2815495)   &  1.997207 (SD: 0.2815495)         \\
$\mu_2$    &  3.995872 (SD: 0.4210236)  &  3.995872 (SD: 0.4210236)        \\
$\sigma^2_1$  & 3.923179 (SD: 0.7913848)  &    4.086645 (SD: 0.8243592)       \\
$\sigma^2_2$  &  8.821342 (SD: 1.775198)
   &      9.188898 (SD: 1.849164)        \\
$\sigma_{12}$ &2.945679 (SD: 0.9464634)  &     3.068416 (SD: 0.9858993)      \\

$\lambda_{11}$   & 1.999832 (SD: 0.2306271)      & 2.040645 (SD: 0.2353338)         \\
$\lambda_{22}$   & 4.998042 (SD: 0.5803669)
           & 5.100043 (SD: 0.5922111)\\
$\lambda_{12}$   &   1.002067 (SD: 0.2672165)    &  1.022517 (SD: 0.2726699)            \\
\midrule
\multicolumn{4}{l}{\textbf{n = 200}} \\ \midrule
$\mu_1$    &  2.001589 (SD: 0.1427427)     &       2.001589 (SD: 0.1427427)     \\
$\mu_2$       & 3.999514 (SD: 0.2126366)   &     3.999514 (SD: 0.2126366)  
\\
$\sigma^2_1$  & 3.979203 (SD: 0.3944584)  &            4.019397 (SD: 0.3984429) \\
$\sigma^2_2$  &  8.944955 (SD: 0.8901909)   &  9.035308 (SD: 0.8991827)           \\
$\sigma_{12}$   & 2.975505 (SD: 0.4695272) &    3.005561 (SD: 0.4742699)         \\

$\lambda_{11}$   & 1.998426 (SD: 0.1153766)      &    2.008469 (SD: 0.1159564)       &    \\
$\lambda_{22}$   &  4.996518 (SD: 0.2876331)          &5.021626 (SD: 0.2890785) \\
$\lambda_{12}$   &  0.9984405 (SD: 0.1352488)     &  1.003458 (SD: 0.1359284)            &        \\
\midrule
\multicolumn{4}{l}{\textbf{n = 500}} \\ \midrule
$\mu_1$   &2.000842 (SD: 0.09020568)       &    2.000842 (SD: 0.09020568)        \\
$\mu_2$       &  4.00109 (SD: 0.1340408)  &4.00109 (SD: 0.1340408)           \\

$\sigma^2_1$  & 3.993712 (SD: 0.2534548)  &      4.009751 (SD: 0.2544727)      \\
$\sigma^2_2$  & 8.982588 (SD: 0.5689367)    &   9.018663 (SD: 0.5712216)   \\
$\sigma_{12}$   &2.996848 (SD: 0.3020145)  &   3.008884 (SD: 0.3032274)        \\

$\lambda_{11}$   &  1.998547 (SD: 0.07260435)     &   2.002552 (SD: 0.07274985)        &    \\
$\lambda_{22}$   & 5.000131 (SD: 0.1804525)           & 5.010151 (SD: 0.1808142)\\
$\lambda_{12}$   &  0.9986841 (SD: 0.08520831)     & 1.000685 (SD: 0.08537907)             &        \\
\bottomrule
\end{tabular}
\end{table}

%%%%%%

%%%%
\begin{table}[H]
\centering
\caption{Comparison of maximum likelihood and Bayesian estimations in Simulation 3}
\label{sim3}
\begin{longtable}{@{}lccc@{}}
\toprule
\textbf{Parameter} & \textbf{MLE} & \textbf{Bayesian}  \\ \midrule\multicolumn{4}{l}{\textbf{n = 25}} \\ \midrule
$\mu_1$       &   2.001857 (SD: 0.1996053)   &  2.001857 (SD: 0.1996053)     &                        \\
              &             &                                          \\
$\mu_2$       &     4.000511 (SD: 0.3989019)     &   4.000511 (SD: 0.3989019)    &                        \\
              &             &                                          \\
$\mu_3$       &  6.00609 (SD: 0.5932688)        &  6.00609 (SD: 0.5932688)          &                        \\
              &             &                                          \\
$\sigma^2_1$  & 0.9618643 (SD: 0.2764356)     &   1.093028 (SD: 0.3141314)          &                        \\
              &             &                                          \\

$\sigma^2_2$  &   3.851198 (SD: 1.113989)      &     4.376362 (SD: 1.265897)         &                        \\
              &             &                                          \\
$\sigma^2_{3}$  &   8.655308 (SD: 2.536946)  &      9.835577 (SD: 2.882893)       &                        \\
              &             &                                          \\
             
$\sigma_{12}$  & 1.348785 (SD: 0.4805106)  &        1.53271 (SD: 0.5460348)   &                        \\
              &             &                                         \\
$\sigma_{13}$  &  0.5855684 (SD: 0.5996525) &         0.6654186 (SD: 0.6814233)      &                        \\
              &             &                                          \\
$\sigma_{23}$  &    1.449337 (SD: 1.20976)  &          1.646974 (SD: 1.374727)    &                        \\
              &             &                                          \\

$\lambda_{11}$   & 2.004908 (SD: 0.3292506)     &      2.11786 (SD: 0.3478)                                  \\
              &             &                                          \\
$\lambda_{12}$   & 1.002803 (SD: 0.3870634)     &   1.059299 (SD: 0.4088698)                                  \\
              &             &                                          \\
$\lambda_{13}$   &    1.002417 (SD: 0.3094444)      &   1.058891 (SD: 0.3268779)                               \\
              &             &                                          \\
$\lambda_{22}$   &   5.010742 (SD: 0.8130529)         &   5.293038 (SD: 0.8588587)                                 \\
              &             &                                          \\
$\lambda_{23}$   &   2.004597 (SD: 0.5016901)         &   2.117532 (SD: 0.5299543)              &                        \\
              &             &                                          \\
$\lambda_{33}$   &  2.999371 (SD: 0.4971395)        &    3.16835 (SD: 0.5251473)            &                        \\
              &             &                                          \\ 
\bottomrule
\end{longtable}
\end{table}  

\begin{table}[H]
\centering
\caption{Comparison of maximum likelihood and Bayesian estimations in Simulation 3 (continued)}
\label{sim3-continued}
\begin{longtable}{@{}lccc@{}}
\toprule
\textbf{Parameter} & \textbf{MLE} & \textbf{Bayesian}  \\ \midrule\multicolumn{4}{l}{\textbf{n = 50}} \\ \midrule
$\mu_1$       & 2.001289 (SD: 0.1414633)    & 2.001289 (SD: 0.1414633)       &                        \\
              &             &                                         \\
$\mu_2$       &     4.006495 (SD: 0.2821739)     &   4.006495 (SD: 0.2821739)      &                        \\
              &             &                                         \\
$\mu_3$       &  6.009829 (SD: 0.4182362)       &    6.009829 (SD: 0.4182362)       &                        \\
              &             &                                          \\
$\sigma^2_1$  &   0.9788372 (SD: 0.1987836)  &    1.041316 (SD: 0.2114719)        &                        \\
              &             &                                          \\

$\sigma^2_2$  &     3.91668 (SD: 0.7861622)    &  4.16668 (SD: 0.8363427)              &                        \\
              &             &                                          \\
$\sigma^2_{3}$  &  8.81281 (SD: 1.784628)
   &   9.37533 (SD: 1.89854)        &                        \\
              &             &                                          \\
             
$\sigma_{12}$  &  1.371165 (SD: 0.3410975)          &  1.458686 (SD: 0.3628697)   &                 \\
              &             &                                          \\
$\sigma_{13}$  &  0.5826238 (SD: 0.4231293)          &        0.6198126 (SD: 0.4501376)           \\
              &             &                                          \\
$\sigma_{23}$  &   1.456099 (SD: 0.8677798)    & 1.549041 (SD: 0.9231701) &                    \\
              &             &                                          \\

$\lambda_{11}$   &  1.999239 (SD: 0.2297266)    &      2.054012 (SD: 0.2360205)                                  \\
              &             &                                          \\
$\lambda_{12}$   &  0.9989928 (SD: 0.2741133)    &   1.026362 (SD: 0.2816232)                             \\
              &             &                                          \\
$\lambda_{13}$   &   0.9996421 (SD: 0.2188311)       &    1.02703 (SD: 0.2248265)                             \\
              &             &                                          \\
$\lambda_{22}$   &  5.000366 (SD: 0.5758727)          &   5.137362 (SD: 0.59165)                                  \\
              &             &                                          \\
$\lambda_{23}$   &  1.999313 (SD: 0.354906)          &     2.054089 (SD: 0.3646294)                                   \\
              &             &                                          \\
$\lambda_{33}$   &    2.996561 (SD: 0.3473413)      &    3.078658 (SD: 0.3568575)                         \\
              &             &                                         \\ 
\bottomrule
\end{longtable}
\end{table}

%%%%
\begin{table}[H]
\centering
\caption{Comparison of maximum likelihood and Bayesian estimations in Simulation 3 (continued)}
\label{sim3-continued200}
\begin{longtable}{@{}lccc@{}}
\toprule
\textbf{Parameter} & \textbf{MLE} & \textbf{Bayesian}  \\ \midrule\multicolumn{4}{l}{\textbf{n = 200}} \\ \midrule
$\mu_1$       & 1.999994 (SD: 0.07056982) & 1.999994 (SD: 0.07056982) \\ & & \\
$\mu_2$       & 4.000806 (SD: 0.142892) & 4.000806 (SD: 0.142892) \\ & & \\
$\mu_3$       & 6.003733 (SD: 0.2112991) & 6.003733 (SD: 0.2112991) \\ & & \\

$\sigma^2_1$  & 0.9941807 (SD: 0.09982312) & 1.009321 (SD: 0.1013433) \\ & & \\
$\sigma^2_2$  & 3.981968  (SD: 0.3995839) & 4.042607 (SD: 0.4056689) \\ & & \\
$\sigma^2_{3}$  & 8.967562  (SD: 0.8961266) & 9.104123 (SD: 0.9097732) \\ & & \\
             
$\sigma_{12}$  & 1.392808 (SD: 0.1736081) & 1.414019 (SD: 0.1762519) \\ & & \\
$\sigma_{13}$  & 0.5962916  (SD: 0.2150788) & 0.6053722 (SD: 0.2183541) \\ & & \\
$\sigma_{23}$  & 1.490207  (SD: 0.436213) & 1.5129 (SD: 0.4428558) \\ & & \\

$\lambda_{11}$   &  1.999746 (SD: 0.1159753)    &   2.013167 (SD: 0.1167536)                                     \\
              &             &                                          \\
$\lambda_{12}$   & 0.9999034 (SD: 0.1359524)     &     1.006614 (SD: 0.1368648)                                  \\
              &             &                                          \\
$\lambda_{13}$   &    0.9997189 (SD: 0.1086336)      &    1.006428 (SD: 0.1093627)                                   \\
              &             &                                          \\
$\lambda_{22}$   &   4.997789 (SD: 0.2901505)         &   5.031331 (SD: 0.2920978)                                     \\
              &             &                                          \\
$\lambda_{23}$   &    1.998484 (SD: 0.1780972)        &  2.011896 (SD: 0.1792925)                                       \\
              &             &                                          \\
$\lambda_{33}$   &  2.999204 (SD: 0.1738039)        &    3.019332 (SD: 0.1749704)                                  \\
              &             &                   \\
\bottomrule

\end{longtable}

\end{table}
%%%%%
%%%%
\begin{table}[H]
\centering
\caption{Comparison of maximum likelihood and Bayesian estimations in Simulation 3 (continued)}
\label{sim3-continued500}
\begin{longtable}{@{}lccc@{}}
\toprule
\textbf{Parameter} & \textbf{MLE} & \textbf{Bayesian}  \\ \midrule\multicolumn{4}{l}{\textbf{n = 500}} \\ \midrule
$\mu_1$       & 1.999996 (SD: 0.04464038)    &   1.999996 (SD: 0.04464038)                           \\
              &             &                                         \\
$\mu_2$      &  3.999947 (SD: 0.08934335)      &   3.999947 (SD: 0.08934335)     &                        \\
              &             &                                          \\
$\mu_3$       & 5.999125 (SD: 0.1331876)        &    5.999125 (SD: 0.1331876)       &                        \\
              &             &                                          \\
$\sigma^2_1$  & 0.9984463 (SD: 0.0627888)     &    1.004473 (SD: 0.06316781)       &                        \\
              &             &                                          \\

$\sigma^2_2$  &     3.992415 (SD: 0.2512253)  &                 4.016514 (SD: 0.2527417)  &                   \\
              &             &                                          \\
$\sigma^2_{3}$  &  8.990107 (SD: 0.5671697)   &     9.044374 (SD: 0.5705933)                               \\
              &             &                                          \\
             
$\sigma_{12}$  &    1.397712 (SD: 0.1085339)         &  1.406149 (SD: 0.109189)                  \\
              &             &                                          \\
$\sigma_{13}$  & 0.5989173 (SD: 0.136427)           &      0.6025325 (SD: 0.1372505)           \\
              &             &                                          \\
$\sigma_{23}$  &     1.499177 (SD: 0.2751862)  &   1.508226 (SD: 0.2768472)           \\
              &             &                                          \\

$\lambda_{11}$   & 1.999446 (SD: 0.0730402)     &  2.004792 (SD: 0.07323549)               &                        \\
              &             &                                          \\
$\lambda_{12}$   & 0.9997517 (SD: 0.08515412)     & 1.002425 (SD: 0.08538181)              &                        \\
              &             &                                          \\
$\lambda_{13}$   &   0.999939 (SD: 0.06844347)       &      1.002613 (SD: 0.06862647)        &                        \\
              &             &                                          \\
$\lambda_{22}$   &  4.999755 (SD: 0.1830031)          &   5.013123 (SD: 0.1834924)              &                        \\
              &             &                                          \\
$\lambda_{23}$   &  1.999826 (SD: 0.112514)          &    2.005173 (SD: 0.1128148)             &                        \\
              &             &                                          \\
$\lambda_{33}$   &    3.000691 (SD: 0.1095989)      &      3.008714 (SD: 0.109892)          &                        \\
              &             &                                         \\ 
\bottomrule
\end{longtable}
\end{table}

\section{Real data}
{\bf Example 1. (Medical dataset)}
The data presented in Table \ref{ex1} represent the range of pulse rate over a day ($X_1$), the range of systolic blood pressure over the same day ($X_2$), and the range of diastolic blood pressure over the same day ($X_3$). These observations were gathered from a sample of 59 patients, each suffering from various illnesses, out of a yearly hospitalized population of 3000. This data was used by Gil et al. (2007). 

\noindent Despite $\Theta^{x_3}_{i1}$ being slightly deviant from the normality assumption, overall, having p-values $0.14$, and $0.08$ using Mardia's test or the Shapiro-Wilk test (see Korkmaz et al. 2014) indicates that there is no significant evidence against multivariate normality of $\bm{\Theta}_{i1}=(\Theta^{x_1}_{i1}, \Theta^{x_2}_{i1}, \Theta^{x_3}_{i1})^\top$. As a means to verify whether 
$$\bm{\Theta}_{i2}=\begin{bmatrix}
\Theta^{x_1}_{i2} & \Theta^{x_1 x_2}_{i2} & \Theta^{x_1 x_3}_{i2}\\
\Theta^{x_1 x_2}_{i2} & \Theta^{x_2}_{i2} & \Theta^{x_2 X_3}_{i2}\\\Theta^{x_1 x_3}_{i2} & \Theta^{x_2 X_3}_{i2} & \Theta^{X_3}_{i2}
\end{bmatrix} $$ is consistent with being drawn from a Wishart distribution, we follow Algorithm \ref{alg:gof_wishart} with $n=59$, $p=3$, degrees of freedom, $df=n-p+1=57$, and ML estimator of $\Lambda$ from Table \ref{tab:parameter-estimates}.

\begin{algorithm}
\caption{Goodness-of-Fit Test for Wishart Distribution}
\label{alg:gof_wishart}

\begin{algorithmic}[1]
\STATE \textbf{Input:} Observed matrices $\{O_1, O_2, ..., O_n\}$, Wishart parameters $df$ and $\hat{\Lambda}$
\STATE \textbf{Output:} Test statistic and p-value

\STATE Simulate $n$ Wishart-distributed matrices $\{S_1, S_2, ..., S_n\}$ with parameters $df$ and $\hat{\Lambda}$
\FOR{$i = 1$ to $n$}
    \STATE Compute the sample covariance matrix $C_i$ of $O_i$
    \STATE Reshape $C_i$ into a vector $v_i$
\ENDFOR
\STATE Reshape each $S_i$ into a vector $u_i$
\STATE Perform a statistical test to compare the distributions of $\{v_1, v_2, ..., v_n\}$ and $\{u_1, u_2, ..., u_n\}$ (Chi-squared test)
\STATE Compute the test statistic and p-value
\RETURN p-value
\end{algorithmic}
\end{algorithm}
Having returned p-value=0.2 does not reject the assumption that $\bm{\Theta}_{i2}$ is from a Wishart distribution.
\newpage

\begin{table}[H]
\centering
\caption{Data on the ranges of pulse rate ($X_1$), systolic ($X_2$), and diastolic ($X_3$) blood pressure}
\begin{tabular}{cccccc}
\toprule
$X_1$ & $X_2$ & $X_3$ & $X_1$ & $X_2$ & $X_3$ \\
\midrule
58--90 & 118--173 & 63--102 & 52--78 & 119--212 & 47--93 \\
47--68 & 104--161 & 71--118 & 55--84 & 122--178 & 73--105 \\
32--114 & 131--186 & 58--113 & 61--101 & 127--189 & 74--125 \\
61--110 & 105--157 & 62--118 & 65--92 & 113--213 & 52--112 \\
62--89 & 120--179 & 59--94 & 38--66 & 141--205 & 69--133 \\
63--119 & 101--194 & 48--116 & 48--73 & 99--169 & 53--109 \\
51--95 & 109--174 & 60--119 & 59--98 & 126--191 & 60--98 \\
49--78 & 128--210 & 76--125 & 59--87 & 99--201 & 55--121 \\
43--67 & 94--145 & 47--104 & 49--82 & 88--221 & 37--94 \\
55--102 & 148--201 & 88--130 & 48--77 & 113--183 & 55--85 \\
64--107 & 111--192 & 52--96 & 56--133 & 94--176 & 56--121 \\
54--84 & 116--201 & 74--133 & 37--75 & 102--156 & 50--94 \\
47--95 & 102--167 & 39--84 & 61--94 & 103--159 & 52--95 \\
56--90 & 104--161 & 55--98 & 44--110 & 102--185 & 63--118 \\
44--108 & 106--167 & 45--95 & 46--83 & 111--199 & 57--113 \\
63--109 & 112--162 & 62--116 & 52--98 & 130--180 & 64--121 \\
62--95 & 136--201 & 67--122 & 56--84 & 103--161 & 55--97 \\
48--107 & 90--177 & 52--104 & 54--92 & 125--192 & 59--101 \\
26--109 & 116--168 & 58--109 & 53--120 & 97--182 & 54--104 \\
61--108 & 98--157 & 50--111 & 49--88 & 124--226 & 57--101 \\
54--78 & 98--160 & 47--108 & 75--124 & 120--180 & 59--90 \\
53--103 & 97--154 & 60--107 & 58--99 & 100--161 & 54--104 \\
47--86 & 87--150 & 47--86 & 59--78 & 159--214 & 99--127 \\
70--132 & 141--256 & 77--158 & 55--89 & 138--221 & 70--118 \\
63--115 & 108--147 & 62--107 & 55--80 & 87--152 & 50--95 \\
47--83 & 115--196 & 65--117 & 70--105 & 120--188 & 53--105 \\
56--103 & 99--172 & 42--86 & 40--80 & 95--166 & 54--100 \\
71--121 & 113--176 & 57--95 & 56--97 & 92--173 & 45--107 \\
68--91 & 114--186 & 46--103 & 37--86 & 83--140 & 45--91 \\
62--100 & 145--210 & 100--136 \\
\bottomrule
\end{tabular} \label{ex1}
\end{table}

%\begin{figure}
%\centering
%\begin{subfigure}[b]{0.45\textwidth}
 % \centering
  %\includegraphics[width=\textwidth]{pretheta1a}
  %\caption{Histogram of $\Theta^{x_1}_{i1}$, $\Theta^{x_2}_{i1}$, and $\Theta^{x_3}_{i1}$}
  %\label{fig:pretheta1a}
%\end{subfigure}
%\begin{subfigure}[b]{0.45\textwidth}
 % \centering
 % \includegraphics[width=\textwidth]{posttheta1a}
 % \caption{Histogram of $\Theta^{*x_1}_{i1}$, $\Theta^{*x_2}_{i1}$, and $\Theta^{*x_3}_{i1}$}
 % \label{fig:posttheta1a}
%\end{subfigure}

%\begin{subfigure}[b]{0.45\textwidth}
 % \centering
 % \includegraphics[width=\textwidth]{pretheta2a}
  %\caption{Histogram of $\Theta^{x_1}_{i2}$, $\Theta^{x_2}_{i2}$, and $\Theta^{x_3}_{i2}$}
  %\label{fig:pretheta2a}
%\end{subfigure}
%\begin{subfigure}[b]{0.45\textwidth}
 % \centering
  %\includegraphics[width=\textwidth]{posttheta2a}
  %\caption{Histogram of $\Theta^{*x_1}_{i2}$, $\Theta^{*x_2}_{i2}$, and $\Theta^{*x_3}_{i2}$}
  %\label{fig:posttheta2a}
%\end{subfigure}

%\caption{Histogram of pre and post transformed variables }
%\label{fig:entire-figure}
%\end{figure}

The ML and Bayesian estimators for mean vector $\bm{\mu}$, variance covariance matrix $\bm{\Sigma}$, and scale matrix $\bm{\Lambda}$ are presented in Table \ref{tab:parameter-estimates}.

\begin{table}[H]
\centering
\caption{Parameter Estimates for data in Table \ref{ex1} }
\label{tab:parameter-estimates}
\begin{tabular}{lccc}
\toprule
\textbf{Parameter} & \textbf{MLE} & \textbf{Bayesian} \\
\midrule
$\bm{\mu}$ & $\begin{bmatrix}
    74.5169 \\
    146.7034 \\
    83.4491 \\
\end{bmatrix}$ & $\begin{bmatrix}
    74.5169 \\
    146.7034 \\
    83.4491 \\
\end{bmatrix}$ \\ \midrule
$\bm{\Sigma}$ & $\begin{bmatrix}
   116.08446&   27.03893&   18.16188 \\
27.03893&  329.96711&  149.77729 \\
18.16188&  149.77729&  157.47199\\
\end{bmatrix}
$ & $\begin{bmatrix}
    122.30327 &  28.48744&   19.13484\\
28.48744 & 347.64392&  157.80107\\
19.13484 & 157.80107&  165.90799\\
\end{bmatrix}
$\\ \midrule
$\bm{\Lambda}$ & $\begin{bmatrix}
    2.7849& 4.1862 & 3.033 \\
    4.1862 & 7.5626 & 5.1744 \\
    3.0329 & 5.1744 &3.7496 \\
\end{bmatrix}
$& $\begin{bmatrix}
    2.7883& 4.1912& 3.036\\
4.1912& 7.5716& 5.1806\\
3.0365& 5.1806& 3.754\\
\end{bmatrix}
$\\
\bottomrule
\end{tabular}
\end{table}

{\bf Example 2. (Car dataset)}

The data in Table \ref{tab:cars_data} provides measurements for 8 different car models. These measurements include four variables: $X_1$ represents the price of the car in thousands of euros, $X_2$ denotes the maximum velocity, $X_3$ indicates the acceleration time required to reach a given speed, and $X_4$ represents the cylinder capacity of the car. These variables are utilized as per Billard and Diday (2012).

\begin{table}[htbp]
    \centering
    \caption{Cars Data}
    \label{tab:cars_data}
    \begin{tabular}{lcccc}
        \toprule
        Car Model & $X_1$ & $X_2$ & $X_3$ & $X_4$ \\
        \midrule
        Aston Martin & $[260.5, 460.0]$ & $[298, 306]$ & $[4.7, 5.0]$ & $[5935, 5935]$ \\
        Audi A6 & $[68.2, 140.3]$ & $[216, 250]$ & $[6.7, 9.7]$ & $[1781, 4172]$ \\
        Audi A8 & $[123.8, 171.4]$ & $[232, 250]$ & $[5.4, 10.1]$ & $[2771, 4172]$ \\
        BMW 7 & $[104.9, 276.8]$ & $[228, 240]$ & $[7.0, 8.6]$ & $[2793, 5397]$ \\
        Ferrari & $[240.3, 391.7]$ & $[295, 298]$ & $[4.5, 5.2]$ & $[3586, 5474]$ \\
        Honda NSR & $[205.2, 215.2]$ & $[260, 270]$ & $[5.7, 6.5]$ & $[2977, 3179]$ \\
        Mercedes C & $[55.9, 115.2]$ & $[210, 250]$ & $[5.2, 11.0]$ & $[1998, 3199]$ \\
        Porsche & $[147.7, 246.4]$ & $[280, 305]$ & $[4.2, 5.2]$ & $[3387, 3600]$ \\
        \bottomrule
    \end{tabular}
\end{table}
\noindent One can easily verify that there is no evidence to reject \(\bm{\Theta}_{i1}=(\Theta^{x_1}_{i1}, \Theta^{x_2}_{i1}, \Theta^{x_3}_{i1}, \Theta^{x_4}_{i1})^\top\) as a multivariate normal. Due to the small sample size, checking whether
$$\bm{\Theta}_{i2}=\begin{bmatrix}
\Theta^{x_1}_{i2} & \Theta^{x_1 x_2}_{i2} & \Theta^{x_1 x_3}_{i2} & \Theta^{x_1 x_4}_{i2}\\
\Theta^{x_1 x_2}_{i2} & \Theta^{x_2}_{i2} & \Theta^{x_2 X_3}_{i2} & \Theta^{x_2 x_4}_{i2}\\\Theta^{x_1 x_3}_{i2} & \Theta^{x_2 X_3}_{i2} & \Theta^{X_3}_{i2} & \Theta^{X_3 X_4}_{i2}\\
 \Theta^{X_1 X_4}_{i2} & \Theta^{X_2 X_4}_{i2}& \Theta^{X_3 X_4}_{i2} & \Theta^{X_4}_{i2} 
\end{bmatrix} $$
follows a Wishart distribution is quite challenging. However, we attempt to modify Algorithm \ref{alg:gof_wishart} to incorporate bootstrapping by resampling from the observed matrices with replacement to generate additional samples. This can be seen in Algorithm \ref{alg:gof_wishart_boot}, with $n=8$, $B=100$, $p=4$, and $df=5$.

\noindent The results (p-value = $0.33$) confirms that we can Wishart distribution assumption is not violated. In Table \ref{tab:cars_data} one can find the ML and Bayesian estimation of the parameters in this example.

%%%%%

\begin{algorithm}
\caption{Goodness-of-Fit Test for Wishart Distribution with Bootstrapping}
\label{alg:gof_wishart_boot}
\begin{algorithmic}[1]
\STATE \textbf{Input:} Observed matrices $\{O_1, O_2, ..., O_n\}$, Wishart parameters $df$ and $\hat{\Lambda}$, Bootstrap iterations $B$
\STATE \textbf{Output:} Test statistic and p-value

\FOR{$b = 1$ to $B$} 
    \STATE Randomly sample $n$ matrices with replacement from $\{O_1, O_2, ..., O_n\}$ to obtain $\{O_{b1}, O_{b2}, ..., O_{bn}\}$
    \STATE Simulate $n$ Wishart-distributed matrices $\{S_{b1}, S_{b2}, ..., S_{bn}\}$ with parameters $df$ and $\hat{\Lambda}$
    \FOR{$i = 1$ to $n$}
        \STATE Compute the sample covariance matrix $C_{bi}$ of $O_{bi}$
        \STATE Reshape $C_{bi}$ into a vector $v_{bi}$
    \ENDFOR
    \STATE Reshape each $S_{bi}$ into a vector $u_{bi}$
    \STATE Perform a statistical test to compare the distributions of $\{v_{b1}, v_{b2}, ..., v_{bn}\}$ and $\{u_{b1}, u_{b2}, ..., u_{bn}\}$ (Chi-squared test)
    \STATE Compute the test statistic and p-value for the $b$th bootstrap iteration
\ENDFOR
\STATE Compute the test statistic and p-value based on the distribution of the bootstrap test statistics
\RETURN p-value
\end{algorithmic}
\end{algorithm}

\begin{table}[H]
\tiny{
\centering
\caption{Parameter Estimates for data in Table \ref{tab:cars_data}}
\label{tab:parameter-estimates-car}
\begin{tabularx}{\textwidth}{l *{2}{>{\centering\arraybackslash}X}}
\toprule
\textbf{Parameter} & \textbf{MLE} & \textbf{Bayesian} \\
\midrule
$\bm{\mu}$ & $\begin{bmatrix}
    201.4687  \\
    261.75 \\
    6.5437 \\
    3772.25\\
\end{bmatrix}$ & $\begin{bmatrix}
    201.4687  \\
    261.75 \\
    6.5437 \\
    3772.25\\
\end{bmatrix}$ \\ \midrule
\!\!\!$\bm{\Sigma}$ & $\begin{bmatrix}
   8040.986 & 2268.914& -109.7983&   81444.44\\
  2268.914 &  852.37&  -42.7609&  19997.9\\
  -109.798&   -42.7609&    2.191&    -903.314\\
 81444.44& 19997.906& -903.314& 1002110.312\\
\end{bmatrix}
$ & $\begin{bmatrix}
   16081.973 & 4537.8281&  -219.596&  162888.88\\
4537.828&  1704.75 &  -85.521&   39995.81\\
-219.59&   -85.521&     4.3830&   -1806.628\\
162888.884& 39995.8125& -1806.62812 &2004220.625\\ 
\end{bmatrix}
$\\ \midrule
$\bm{\Lambda}$ & $\begin{bmatrix}
    235.613&  25.751&  2.774&  2222.525 \\
    25.751&   8.38&  0.99&   414.185 \\
    2.774&   0.99&  0.145&   55.38 \\
    2222.525& 414.185& 55.38& 40736.7\\
\end{bmatrix}
$& $\begin{bmatrix}
   269.271&  29.43& 3.17&  2540.02\\
   29.43&   9.576&  1.131&   473.35\\
   3.17&   1.131&  0.165 &    63.3\\
   2540.03& 473.354& 63.3& 46556.228\\
\end{bmatrix}
$\\
\bottomrule
\end{tabularx}
}
\end{table}

\begin{remark}
    As can be seen from Example 2, when dealing with a scenario where the sample size is small ($n=8$) and the dimensionality of the data is relatively large compared to the sample size ($p=4$), it is common to observe significant divergence between the ML and Bayesian estimators. This discrepancy arises due to several factors. Firstly, in general, the limited amount of data can lead to sparse representations in the high-dimensional space, making it challenging for the ML method to accurately estimate parameters. On the other hand, Bayesian estimation offers advantages in this context. By incorporating prior information about the parameters, Bayesian estimation can provide regularization and help stabilize parameter estimates, particularly when the sample size is small. Moreover, Bayesian methods allow for the integration of domain knowledge and uncertainty quantification, which can improve the robustness of the estimates. Overall, this divergence underscores the superiority of Bayesian estimation over MLE in scenarios with small sample sizes and high dimensionality, highlighting the importance of adopting Bayesian approaches when dealing with such data constraints, while ML relies on large sample theory and can have problems in smaller samples.
\end{remark}

\section{Concluding remarks}\label{sec:concluding}
So far, we have assumed that values within intervals are uniformly distributed. Hence, the ML and Bayesian estimators of the parameters $\bm{\mu}$, $\bm{\Sigma}$, and $\bm{\Lambda}$ from Theorems \ref{MLest} and \ref{Bayesests}, can be expressed in terms of $a_{ji}$ and $b_{ji}$ for $i=1, \dots, n$ and $j=1, \dots, p$, using equations (\ref{thetai1}) and (\ref{thetai2}).

\noindent Additionally, there are other cases where we might assume that the internal distribution follows other distributions such as the triangular or Pert distribution. Interested readers are referred to Samadi et al. (2023) for more information.

Note that in all aforementioned cases, we maintain the assumption that $\bm{\Theta}_{1}$ and $\bm{\Theta}_{2}$, the mean and the variance-covariance matrix are multivariate normal and Wishart distributed.

Alternatively, some researchers have recently considered (e.g., Brito and Duarte Silva 2012; Lin et al. 2022; and Bertrand et al. 2023) the skew-normal distribution as a generalization when the normality assumption is not met. In the univariate case, $p=1$, a Box-Cox transformation
\[
\Theta^{*x_j}_{i1} = 
\begin{cases} 
    \frac{(\Theta^{x_1}_{i1})^{\kappa_1} - 1}{\kappa_1} & \text{if } \kappa_1 \neq 0 \\
    \log(\Theta^{x_1}_{i1}) & \text{if } \kappa_1 = 0 
\end{cases}\,,
\]
or a power transformation 
\[
\Theta^{*x_1}_{i2} = 
\begin{cases} 
    (\Theta^{x_1}_{i2})^{\kappa_2} & \text{if } \kappa_2 \neq 0 \\
    \log(\Theta^{x_1}_{i2}) & \text{if } \kappa = 0 
\end{cases}\,,
\]
were suggested by Xu and Qin (2022) to convert $\Theta^{x_1}_{ik}$, for $k=1,2$, and $i=1, \dots, n$, into $\Theta^{*x_1}_{ik}$ (which is normal for $k=1$, and exponential for $k=2$). Parameters $\kappa_k$ for $k=1,2$ vary between $(-3, 3)$, and their optimal values can be found numerically.

In conclusion, this paper comprehensively explored both Bayesian and frequentist approaches for handling multivariate interval-valued data. The study involved investigating some asymptotic properties of the ML estimators and comparing ML methods with Bayesian methods in terms of the risk function associated with $L_2$ loss and entropy losses. Through extensive simulation studies, we demonstrated the performance of the proposed estimators and illustrated their applicability to real interval-valued datasets. Overall, our findings underscore the significance of leveraging both Bayesian and frequentist methodologies in addressing the complexities of multivariate interval-valued data, providing valuable insights for future research and practical applications in various domains.

\noindent {\bf Funding:} This research received no external funding.

\noindent {\bf Conflicts of Interest:} The authors declare no conflict of interest. Both co-authors contributed equally to this research effort.

%%%%%%%%%%%%%%%%
\bibliographystyle{plain}

\end{document}